\documentclass[journal]{IEEEtran}
\usepackage[space]{cite}
\usepackage{amsmath,amssymb,amsfonts}
\usepackage{algorithmic}
\usepackage{graphicx}
\usepackage{subcaption}
\usepackage{textcomp}
\usepackage{multirow}
\usepackage[ruled]{algorithm2e}
\usepackage{mathrsfs} 
\usepackage{soul}

\usepackage[dvipsnames]{xcolor}
\def\BibTeX{{\rm B\kern-.05em{\sc i\kern-.025em b}\kern-.08em
    T\kern-.1667em\lower.7ex\hbox{E}\kern-.125emX}}

\newcommand{\mr}[2]{\multirow{#1}{*}{#2}}

\usepackage{array}
\newcolumntype{x}[1]{>{\centering\arraybackslash\hspace{0pt}}p{#1}}

\newcommand{\probP}{\text{I\kern-0.15em P}}

\DeclareMathOperator*{\argmax}{\arg\!\max}

\newcommand{\changed}[1]{{\color{black}#1}}
\newcommand{\removed}[1]{}

\makeatletter
\newcommand{\subalign}[1]{%
  \vcenter{%
    \Let@ \restore@math@cr \default@tag
    \baselineskip\fontdimen10 \scriptfont\tw@
    \advance\baselineskip\fontdimen12 \scriptfont\tw@
    \lineskip\thr@@\fontdimen8 \scriptfont\thr@@
    \lineskiplimit\lineskip
    \ialign{\hfil$\m@th\scriptstyle##$&$\m@th\scriptstyle{}##$\hfil\crcr
      #1\crcr
    }%
  }%
}
\makeatother
\usepackage[utf8]{inputenc}

\begin{document}

\title{\huge Enhancing Automotive User Experience with Dynamic Service Orchestration for Software Defined Vehicles}


\author{
    \IEEEauthorblockN{Pierre Laclau$^{1,2}$, Stéphane Bonnet$^1$, Bertrand Ducourthial$^1$, Xiaoting Li$^2$ and Trista Lin$^2$}

    \small{$^1$Heudiasyc, CNRS, Université de Technologie de Compiègne, France \{firstname.lastname@utc.fr\}}
    \\
    \small{$^2$Stellantis, Vélizy-Villacoublay, France \{firstname.lastname@stellantis.com\}}
}

\maketitle

\begin{abstract}
    With the increasing demand for dynamic behaviors in automotive use cases, Software Defined Vehicles (SDVs) have emerged as a promising solution by bringing dynamic onboard service management capabilities. While users may request a wide range of services during vehicle operation, background tasks such as cooperative Vehicle-to-Everything (V2X) services can activate on-the-fly in response to real-time road conditions. In this dynamic environment, the efficient allocation of onboard resources becomes a complex challenge, in order to meet mixed-criticality onboard Quality-of-Service (QoS) network requirements while ensuring an optimal user experience (UX). Additionally, the ever-evolving real-time network connectivity and computational availability conditions further complicate the process.
    In this context, first we introduce the concept of Automotive eXperience Integrity Level (AXIL) which expresses a runtime priority for non-safety-critical applications to enable dynamic orchestration. Second, we present a dynamic resource-based onboard service orchestration algorithm that considers real-time in-vehicle and V2X network health, along with onboard resource constraints, to select degraded modes for onboard applications and maximize user experience. Third, we show by simulation that our algorithm produces near-optimal solutions while significantly reducing execution time: it satisfies approximately 90\% of the onboard UX compared to the optimal while reducing the execution time of one to two orders of magnitude compared to an exact solver. With this approach, we aim to enable efficient onboard execution for a UX-focused service orchestration.
\end{abstract}
\noindent\let\thefootnote\relax\footnote{This work was supported by Stellantis under the collaborative CIFRE framework UTC/CNRS/PCA (ANRT contract n°2021/0865) with Heudiasyc.}

\begin{IEEEkeywords}
    Software Defined Vehicle (SDV), User Experience (UX), Service Oriented Architecture (SOA), Optimization, Network Aware Orchestration, In Vehicle Network (IVN).
\end{IEEEkeywords}

\section{Introduction}

In the last few years, automotive manufacturers have started to develop a range of complex features such as dedicated infotainment app stores, automated driving, and Vehicle-to-Everything (V2X) services. This trend invites automotive manufacturers to continuously deliver these features through seamless Over-the-Air (OTA) updates, which contributes to longer lasting vehicles in the hope to bring the industry closer to sustainable transportation systems. As a result, future vehicles may resemble '\textit{Smartphones on Wheels}' where users can dynamically request services throughout the vehicle lifetime while automakers continuously integrate background services such as Advanced Driver Assistance Systems (ADAS), cooperative multi-vehicle services, smart grid, and more \cite{haeberleSoftwarization2020,huEnergy2018}.

To support this paradigm shift, the automotive industry is undergoing a rapid transformation toward Software Defined Vehicles (SDV) \cite{zhao_identification_2022}. Previously, Electric and Electronic (E/E) architectures were hardware-defined by integrating many single-function Electronic Control Units (ECU) into domain-centric networks. However, the increasing number of ECUs as well as more demanding network and OTA requirements are reaching the limits of current architectures \cite{bandurMaking2021}. These changes are motivating a global shift towards Zonal Oriented Architectures (ZOA), where fewer High Performance Computing (HPC) ECUs are expected to group and manage multiple heterogeneous functions. ZOA allows for a centralized and therefore cost-effective reservation of computational resources for future updates and features. In addition, Ethernet facilitates service-oriented communications and increases bandwidth \cite{zengInVehicle2016}. Together, these efforts toward centralization using fewer and virtualized ECUs can help reduce resource usage such as embedded hardware needs and runtime energy consumption.

This new hardware approach must be controlled by a software stack capable of applying OTA updates and dynamically switching the onboard software context to provide relevant applications. A common solution is to deploy a Service Oriented Architecture (SOA) which handles high-level communications through publish-subscribe protocols and the orchestration of services based on user requests, vehicle context, and available updates, allowing for dynamic reconfigurations of software and network resources \cite{rumezOverview2020b}. However, SOA does not address the guarantee of Quality of Service (QoS) constraints such as hard real-time latency and jitter for safety-critical flows. Hence, the onboard architecture also relies on an embedded infrastructure based on technologies such as Software Defined Networking (SDN) and Time Sensitive Networking (TSN) that can be used to handle dynamic routing and hard real-time scheduling of network resources on a per-flow basis \cite{lin_adaptive_2021}.

While reconfigurations are feasible, the simultaneous activation of the rising number of applications from app stores and updates could face limitations due to onboard constraints, potentially hindering on the user experience (UX). The onboard constrained resources, coupled with strict automotive requirements \cite{askaripoorArchitecture2022a}, may impose limitations on the sets of possible combinations of concurrently active applications. Hence, the vehicle may need to prioritize and schedule specific app combinations within its physical capacities. In certain scenarios, it could become necessary to temporarily degrade the quality of service allocated to certain applications or even shutdown some of them, ideally depending on UX impacts~\cite{dai_quality_2016}.

However, there is currently no method for selecting which applications to degrade or shutdown when the physical capabilities are not sufficient to allocate all desired services. Besides, the automotive industry must ensure that a user will have an enjoyable experience as defined by contract-based Service Level Agreements (SLAs) established at the time of vehicle purchase \cite{liAutomated2021}. Depending on the customer's vehicle model, profile, and subscriptions, automakers may define different runtime priorities to adapt to the user expectations (i.e. higher-level SLAs should guarantee higher QoS). Currently, the automotive industry does not have a methodology to dynamically adapt the vehicle's QoS depending on the context.

Without any orchestration methodology, one solution would be to statically allocate the maximum network requirements for each application under the worst network conditions. However, this introduces unnecessary overhead and constrains automakers to include fewer services running in the vehicle.

We propose a different approach by defining a concept called “Automotive eXperience Integrity Level” (AXIL) that takes inspiration from the well-known Automotive Safety Integrity Level (ASIL) standard \cite{frigerioComponentLevel2019}. While ASIL helps engineers optimize the “safety to cost ratio” of onboard features, AXIL aims to help shape the automotive digital offering by introducing a measurable “user experience to onboard resources ratio” for each application. We propose to let each application developer define several runtime modes that require different hardware and network resources with varying levels of functionality. Intermediate modes can be seen as \textit{degraded} modes with reduced functionality. By associating an AXIL rating to each runtime mode, the onboard software infrastructure will then be capable of activating the relevant runtime modes to meet all onboard constraints while maximizing the overall onboard experience, thereby optimizing resource usage. With this approach, the onboard orchestrator may find that slightly degrading some applications to make room for more applications improves the overall perceived~experience for passengers.

Our key contributions are threefold. First, we introduce our AXIL metric coupled with multiple runtime modes for each application to assess the perceived user experience of each feature, facilitating effective orchestration. Second, we introduce an easy-to-represent model of the onboard resource requirements and availability. Based on this model, we seek an efficient algorithm to select modes that maximize the overall user experience while staying within resource limits. This led us to our third and main contribution, where we design a specialized heuristic that strikes a balance between performance and speed, enabling real-time onboard calculations. We present our implementation in a simulated environment to compare it with other approaches. Hence, this paper combines these elements to form a system that dynamically manages onboard services despite runtime changes such as vehicle connectivity.

\begin{figure}[t!]
    \centering
    \includegraphics[width=\linewidth]{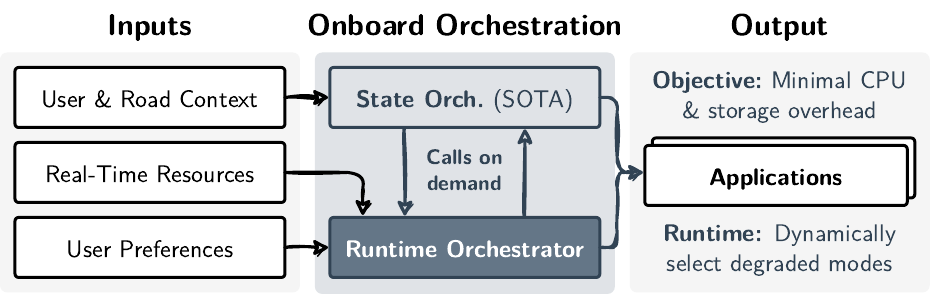}
    \caption{Block diagram of the onboard orchestration platform. Our goal is to orchestrate the selection of runtime modes (including shutdown) for each application to remain within the available onboard, V2X, and edge offloading resources while maximizing the user experience (defined by our AXIL rating).}
    \label{fig:arch}
\end{figure}

Figure \ref{fig:arch} summarizes our approach by identifying the components and external interfaces of our orchestration platform. Our algorithm is invoked on onboard vehicle context or resource availability changes for a continuous user experience.

The remaining sections of this paper are organized as follows. Section~\ref{sec:background} introduces the AXIL metric and the concept of runtime modes. Section~\ref{sec:problem} presents the onboard resource requirements model. Section~\ref{sec:methodology} first discusses potential state-of-the-art optimization algorithms before delving into our proposed heuristic. Sections~\ref{sec:simulation} and~\ref{sec:results} describe our simulation environment and compare the performance of each algorithm.

\section{Background} \label{sec:background}

Traditionally, vehicles are manufactured with a pre-defined static allocation of resources for onboard applications. To ensure a safe and consistent state across the entire vehicle software stack, engineers also define rule-based local state management strategies for each ECU and software layer to account for unexpected events such as link failures, invalid software states, or unexpected packet receptions. Based on the surrounding local state and events, these strategies apply degraded modes on each application with reduced functionality \cite{kugeleServiceOrientation2017}. However, with the arrival of dynamic use cases and updates with thousands of applications, it may become too difficult to control the exact combinations of active applications requested by the user (e.g. infotainment and ADAS features), vehicle (e.g. health monitoring and data collection), or road context (e.g. cooperative V2X services and edge offloading) \cite{schindewolfComparison2022a}. In this context, the traditional rule-based approach lacks the adaptability required to handle unexpected or novel situations, making it inadequate~for~addressing~the intricate and rapidly changing network traffic requirements.

The industry is currently shifting toward dynamic onboard service orchestration using the new capabilities of SOA to enable more efficient resource usage, reduced architecture complexity, and less engineering efforts for state management \cite{frtunikjRunTime2014}. This new paradigm enables vehicle-wide monitoring \cite{liao_cooperative_2022} for centralized and context-aware global orchestration. With the help of reconfigurable technologies such as TSN and SDN for the in-vehicle network, it is now possible to dynamically change the behavior and resource allocations of onboard software components through reconfigurations \cite{shiRecent2023,halbaRobust2018}. Furthermore, several research initiatives propose to allocate network and computing resources using atomic vehicle-wide reconfiguration strategies to guarantee safe transitions even while driving \cite{hackelSoftwareDefined2019b,hauggSimulationbased,haeberleSoftwarization2020}. Hence, while it may be feasible to automate the generation of local state management rules, these methods rely on complex and intensive engineering processes. A new challenge then arises: how can we build a centralized and autonomous state management stack to enable these future dynamic orchestration mechanisms?

Vehicles can execute applications of two categories. First, \textit{safety-critical} (SC) services require Quality-of-Service (QoS) guarantees such as real-time computing and time-sensitive network flows with pre-defined latency, jitter, and isolation requirements. They also require deterministic behavior across the entire vehicle to ensure a safe and reliable execution for any external context. Second, \textit{best-effort} (BE) applications are usually user-focused and offer onboard features to enhance user experience. While some of these applications can require resources with QoS allocations (e.g. video network flow for a video streaming app), others can seamlessly run with a minimum bandwidth requirement without being affected by higher latencies and jitter caused by other apps \cite{pengSurvey2023,kulzerNovel2020}.

Hence, the vehicle could dynamically allocate resources for best-effort applications depending on the available resources.

Our previous work \cite{laclauPredictive2023} presented a methodology to pre-compute configurations for safety-critical applications. It consists in generating resource allocations to guarantee all QoS network requirements based on a fixed onboard E/E topology description. As configuration scheduling takes time and safety certifications require deterministic and predictable vehicle states, our work pre-computes these configurations offboard by predicting the combinations of active applications onboard, then pre-downloads them to the vehicle. With this method, automakers have the time to apply all required safety validation procedures and the vehicle can safely switch between configurations for different execution contexts.

However, with the rising number of non-safety-critical and increasingly dynamic applications, future use cases may include thousands of apps which could reduce the effectiveness of vehicle state prediction. The method of \cite{laclauPredictive2023} cannot guarantee that a valid configuration has been pre-calculated~for every combination of active applications. Additionally, resource availability can change at runtime due to (1) the dynamic activation of safety-critical (SC) apps with reserved network and computing resources thereby restraining best-effort (BE)~applications, and (2) outside conditions such as V2X network congestion or 4G/5G connectivity losses which can severely impact the functionality of applications \cite{kulzerNovel2020}. 

Hence, these BE applications must adapt to the ever-changing remaining unused resources which cannot be fully predicted by methods such as \cite{laclauPredictive2023} focused on SC services. Therefore, this current work addresses these limitations by extending \cite{laclauPredictive2023} with dynamic suboptimal onboard decisions.

Research has already widely explored dynamic resource-based service orchestration strategies, notably in the cloud \cite{ranjanCloud2015,tosattoContainerBased2015a,jiangChallenges2018} and more recently automotive \cite{nayakAutomotive2023,schindewolfResilient2022} industries with bridges between the two for edge and fog computing \cite{nobreVehicular2019,raniVehicular2021}. These initiatives can allocate services to computing nodes (i.e. server or ECU) depending on node-specific peripherals or resource requirements defined by each application developer such as CPU, RAM, GPUs, actuators, and network usage.

However, cloud-native solutions are loosely constrained and therefore only consider simple metrics such as CPU and memory, as opposed to the highly embedded automotive environment with real-time traffic shaping, V2X network constraints, and more. To the best of our knowledge, while some automotive contributions have enhanced the resource models \cite{schindewolfResilient2022}, no research considers the selection of degraded or runtime modes as an additional degree of freedom for more orchestration granularity. For instance, the vehicle could temporarily lower video streaming quality on network congestion events or even activate more applications with newly freed resources.

To the best of our knowledge, there is no existing approach that connects user experience to dynamic vehicle orchestration. The primary technical gaps in existing works include a lack of adaptability in handling dynamic, user-initiated application demands, insufficient consideration of real-time network conditions, and the absence of a framework connecting user experience to dynamic service orchestration. These gaps led to the design of our methodology to optimize onboard user experience through dynamic, context-aware resource allocation.

Therefore, we propose to take inspiration from the well-adopted ASIL standard and extend it for non-safety-critical applications. While ASIL guides engineering design choices to dimension the hardware based on each feature's safety requirements, our proposed AXIL metric can facilitate automated orchestration algorithms for optional user-oriented features.

When allocating applications to specific software contexts or ECUs, automakers can either pre-define them or implement one of many existing scheduling strategies \cite{schindewolfResilient2022}. Here, we propose to add an onboard runtime selector alongside this orchestration step to dynamically apply degraded modes onboard thereby maximizing UX. In this work, we aim to define a light-weight algorithm to select runtime modes, which can then be triggered onboard on context or resource availability changes.

We believe that the unique challenges posed by an environment with thousands of applications are not addressed by existing research. By defining an automated centralized orchestrator with these objectives, we add a global optimization layer of runtime mode selection instead of local rule-based state management, thereby optimizing the onboard user experience and resource usage. Our work aims to bridge this research gap by optimizing the runtime conditions of services in real time to support (1) a large number of applications, (2) maximized user experience, and (3) minimal onboard~overhead.



\section{Problem formulation} \label{sec:problem}

We propose a new resource-based orchestration methodology that allows for granular and real-time runtime mode selection for each application requested by the user or vehicle. In this section, we describe the problem by (A) setting our initial assumptions, (B) defining an app store with runtime modes and dependencies, (C) proposing a data structure to represent the vehicle state, (D) defining the optimization objective, and (E) identifying the NP-Hard nature of the presented problem.

\subsection{Assumptions and goals}

Our approach is based on the assumption that each application has already been assigned to a host software context or ECU, either manually or with the help of an automated onboard orchestrator. We believe this to be a reasonable assumption for the coming architectures and use cases, as services are usually highly dependent on hardware capabilities such as infotainment apps in the High Performance Computer (HPC) or external networking services in the connectivity~ECU.

Additionally, safety-critical applications must reserve resources with guaranteed Quality of Service (QoS) requirements. As such, we assume that resources for these apps have already been reserved by another onboard system (out of scope of this work) prior to this work's algorithm decision time. Our algorithm then uses the remaining unallocated resources to be used for its decision (i.e. set as the $R$ vector).

\changed{We also assume that developers provide accurate maximum usage estimates for each resource. They can use statistical, historical, or worst-case analysis. Optionally, the vehicle can enforce bandwidth limitations for each application using network shaping mechanisms such as TSN traffic shapers.}

\removed{Our goal is to define an efficient \mbox{algorithm} capable of real-time onboard decisions to automatically adjust the onboard functionalities when the onboard or external network resources cannot allocate the currently requested resources. The algorithm must maximize the user experience at all times.}

\subsection{AXIL: User Experience Metric}

As one cannot improve what cannot be measured, we propose to define a new metric to assess the user experience of each application. This metric can be defined by each application developer and extended to each degraded runtime mode to tweak the onboard functionality with enhanced granularity. We call this metric AXIL for \textbf{A}utomotive e\textbf{X}perience \textbf{I}ntegrity \textbf{L}evel. AXIL enables the vehicle to select which functionalities to activate for each application. We identify three parameters to assess the priority of a particular feature offered by an app:

\begin{table}[t!]
    \caption{Definition of AXIL. Just like ASIL, AXIL combines three parameters to assess the runtime priority of a service.}
    \centering

    \begin{tabular}{l|l||cccc}
        \multicolumn{1}{x{0.9cm}|}{\mr{2}{$E_1$}} & \multicolumn{1}{x{0.9cm}||}{\multirow{2}{*}{$E_2$}} & \multicolumn{4}{c}{$E_3$}                                                                                                         \\
                                                  &                                                     & \multicolumn{1}{x{1cm}}{Minimal} & \multicolumn{1}{x{1cm}}{Low} & \multicolumn{1}{x{1cm}}{Medium} & \multicolumn{1}{x{1cm}}{High} \\
        \hline
        \mr{4}{Easy}                              & Rare                                                & -                                & -                            & -                               & -                             \\
                                                  & Low                                                 & -                                & -                            & -                               & -                             \\
                                                  & Medium                                              & -                                & -                            & -                               & A                             \\
                                                  & High                                                & -                                & -                            & A                               & B                             \\
        \hline
        \mr{4}{Medium}                            & Rare                                                & -                                & -                            & -                               & -                             \\
                                                  & Low                                                 & -                                & -                            & -                               & A                             \\
                                                  & Medium                                              & -                                & -                            & A                               & B                             \\
                                                  & High                                                & -                                & A                            & B                               & C                             \\
        \hline
        \mr{4}{Difficult}                         & Rare                                                & -                                & -                            & -                               & A                             \\
                                                  & Low                                                 & -                                & -                            & A                               & B                             \\
                                                  & Medium                                              & -                                & A                            & B                               & C                             \\
                                                  & High                                                & A                                & B                            & C                               & D                             \\
    \end{tabular}

    \vspace{1mm}
    \hspace{25mm} \textit{Priority order: - $<$ A $<$ B $<$ C $<$ D }

    \vspace{5mm}

    \begin{tabular}{c|c|c|c}
        Legend $\rightarrow$ & $E_1$                & $E_2$              & $E_3$           \\
        \hline
        \textbf{ASIL}        & Controllability      & \mr{2}{Exposition} & Severity        \\
        \textbf{AXIL}        & Ease of Substitution &                    & Quality of Exp. \\
    \end{tabular}

    \label{tab:axil}
\end{table}

\begin{itemize}
    \item \textbf{Ease of substitution:} How easily an average user can obtain the same service provided by the onboard feature if the corresponding feature is not provided by the vehicle. For instance, optional navigation features could be easily offered by the user's smartphone instead, while cooperative services may require to be executed onboard.
    \item \textbf{Exposition:} Relative assessment of how often an average user may use the feature compared to the feature set. These values compare the frequency of use of each feature to the total number of features in the vehicle.
    \item \textbf{Quality of Experience (QoE):} Relative measurement of user convenience or enjoyment when using the feature. Automakers may conduct UX studies and define guidelines to calculate the corresponding QoE values, although the methodology is out of the scope of this paper.
\end{itemize}

\begin{figure}[t!]
    \centering
    \includegraphics[clip, trim=0cm 0.1cm 0cm 0.2cm, width=\linewidth]{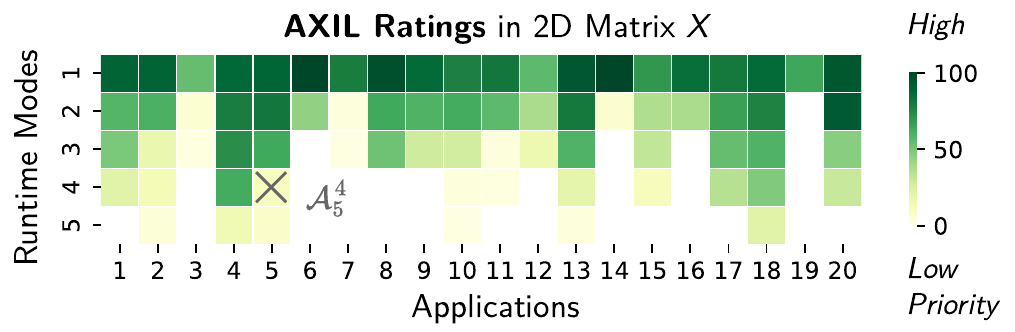}
    \caption{Random problem instance generated with 20 applications and at most 5 modes.  Each cell $(i,j)$ represents mode $j$ of $\mathcal{A}_i$ and is assigned a runtime priority $X_i^j$ (along with a vector of resource requirements $M_i^j$ defined in Section \ref{sec:modes}).}
    \label{fig:instance}
\end{figure}

Then, the original structure defined by ASIL in Table \ref{tab:axil} is an excellent candidate to assign a priority rating based on these three parameters. In the table, AXIL uses the same structure~as ASIL to combine the three previous parameters into one global runtime priority metric. Hence, an AXIL D feature has the highest priority in this definition. Naturally, this concept can be extended with more granular levels and even be dependent on the user profile and preferences. In the rest of this paper, we assume that the AXIL rating can be generalized to any value in $\mathbb{R}^+$ with higher values corresponding to higher priorities. See Figure \ref{fig:instance} for an illustration of a particular instance.

\changed{Potential challenges in defining AXIL ratings include the difficulty of rating each parameter in a harmonized way across services, and evaluating the exposition values. These can be addressed by conducting user surveys or in-vehicle background user observation. We only present the core concept in this work. The actual implementation is left to~the~automakers.}

More broadly, the AXIL rating can be defined as a function of the vehicle state and user profile. For instance, a navigation app may have a higher AXIL rating when the vehicle is in a city compared to a highway. \removed{This outcome could be achieved with existing methods used by web applications, which can ask for regular user ratings of their experience (e.g. call quality for online meeting platforms).} Vehicles could integrate a feedback system where A/B testing methods would study the effectiveness of ratings to gradually converge toward maximized user satisfaction, although not in the scope of this work. Hence, we assume in this work that ratings can evolve dynamically throughout the vehicle lifetime.

\vspace{-1mm}

\subsection{Problem description}

\subsubsection{Applications}
Let $\mathcal{A} = \{\mathcal{A}_1, ..., \mathcal{A}_n\}$ be the set of applications available in the app store, including background services, each with their own independently defined QoS network and other runtime requirements. We partition $\mathcal{A}$ in two disjoint subsets $\mathcal{A}_{\text{SC}} \subseteq \mathcal{A}$ for safety-critical (SC) apps and $\mathcal{A}_{\text{BE}} \subseteq \mathcal{A}$ for best-effort (BE) services such that $n_{\text{SC}} + n_{\text{BE}} = n$ with $n_{\text{SC}} = |\mathcal{A}_{\text{SC}}|$ and $n_{\text{BE}} = |\mathcal{A}_{\text{BE}}|$. We assume that~requirements are provisioned to orchestrate SC apps with methods such as our previous work \cite{laclauPredictive2023}. They may optionally attempt to allocate QoS resources for BE apps without guarantees. This work focuses on allocating BE apps while (1) the vehicle dynamically reserves resources for SC apps and (2) external conditions such as connectivity continuously evolve. It acts as a complementary dynamic layer to our previous work \cite{laclauPredictive2023}.

\subsubsection{Resources}
The vehicle can be modelled as a list of $r$ resources $\mathcal{R}_1, \ldots, \mathcal{R}_r$. The maximal hardware capacity of the vehicle is denoted as a vector of $r$ positive reals $R^{\text{max}}$ such that $R^{\text{max}}[i] \in \mathbb{R}^+$ denotes the maximal capacity of resource $\mathcal{R}_i$. Resources can describe the CPU and memory usages for each ECU, available best-effort bandwidth for each physical network link in each link direction, external system capacity such as V2X and edge computing availability, and more. \changed{We assume that developers provide accurate maximum estimates.}

However, resource availability can dynamically change at runtime, such as when external conditions (e.g. V2X network saturation) or internal events (e.g. resource allocation change due to a change of SC apps thereby impacting BE bandwidth availability) occur. Hence, we define the current BE capacity of a vehicle as a vector $R$ of $r$ elements in  $\mathbb{R}^+$. We have:

$$0 \leq R[i]  \leq R^{\text{max}}[i] \qquad i \in {1,...,r}$$

\subsubsection{Runtime modes} \label{sec:modes}

In this work, we consider that each application $\mathcal{A}_i \in \mathcal{A}_{\text{BE}}$ can independently define $m_i$ runtime (or degraded) modes. Each mode offers a different level of functionalities and resource requirements. For instance, a video streaming application may define multiple video quality options with corresponding network bandwidth usages.
For each application $\mathcal{A}_i$, we denote each mode $j$ as $\mathcal{A}_i^j$ and define $M_i^j$ as its corresponding vector of resource requirements. This is a vector of $r$ positive real elements satisfying:

$$0 \leq M_i^j[k] \leq R^{\text{max}}[k] \qquad k \in {1,...,r}$$

We assume that $\mathcal{A}_i^1$ is the nominal mode with the maximum features, resource requirements, and AXIL rating of $\mathcal{A}_i$. Each following mode provides fewer features using less resources:

$$\forall i = 1, ..., n_{\text{BE}} \quad \forall j, j' = 1, ..., m_i$$
\begin{equation} \label{eq:resources}
    \forall k = 1, ..., r \qquad j<j' \implies M_i^j [k] \ge M_i^{j'} [k]
\end{equation}
$$\text{and} \quad \exists k \in \{1, ..., r\}, \quad M_i^{j} [k] > M_i^{j'} [k]$$

In Figure \ref{fig:instance}, AXIL ratings also follow the previous constraint as higher modes must have lower priorities. Each cell $(i,j)$ is also associated with its vector of resource requirements $M_i^j$.

\subsubsection{Mode dependencies}
In a realistic automotive environment, applications will also have dependencies to other applications. In our context, each runtime mode may define a minimum mode for another application to be active. For instance, the video streaming app may require a minimum display and network activity level managed by their respective background services. Hence, we define a mode dependency graph $D = \{V, E\}$ with nodes $V = \{ \mathcal{A}_i^j \quad \forall 1~\leq~i~\leq~n_{\text{BE}}, \quad 1~\leq~j~\leq~m_i \}$ as a Directed Acyclic Graph (DAG) to avoid cyclic dependencies. There is an oriented edge $(\mathcal{A}_i^j, \mathcal{A}_{i'}^{j'})$ in $E$ if and only if $\mathcal{A}_i^j$ requires~$\mathcal{A}_{i'}^{j'}$.

\subsection{Optimization problem}
The previous definitions describe the problem instance. However, at any time, the vehicle may request to launch any combination of applications $\mathcal{A}_{\text{req}} \subseteq \mathcal{A}_{\text{BE}}$ with $n_{\text{req}} = |\mathcal{A}_{\text{req}}|$.

\subsubsection{Resource model}

The previously defined problem can be arranged in a 3D matrix called $P$ to store the resource requirements (up to $r$) for each runtime mode (up to $m_\text{max} = \max_{i \in \{1, ..., n_{\text{req}}\}} m_i$ defined as the maximum number of modes among the apps $\mathcal{A}_{\text{req}}$) of each application (up to $n_{\text{req}}$):

$$
    P_{i, j, k} =
    \begin{cases}
        M_i^j [k] & \text{if $j \in \{1, ..., m_i\}$} \\
        0         & \text{otherwise}                  \\
    \end{cases}
$$

\subsubsection{User experience}
When the current available resources $R$ cannot allocate all requested applications $\mathcal{A}_{\text{req}}$, the vehicle must select a subset of applications to run. Our approach is to select a mode for each application to maximize the user experience. For each application, each runtime mode can be assigned an AXIL rating to represent a launch priority focused on UX. This information can be stored for all applications in a single 2D matrix $X$ of size $n_{\text{req}}, \, m_{\text{max}}$ with values in $\mathbb{R}^+$.

\subsubsection{Objective function}
Our goal is to select one mode for each application to maximize the user experience while respecting the resource and dependency constraints. The solution can be represented as a vector $C$ of selected runtime modes:

\begin{equation}
    C[i] =
    \begin{cases}
        0 & \text{if $\mathcal{A}_i \in \mathcal{A}_{\text{req}}$ is off} \\
        j & \text{if mode $\mathcal{A}_i^j$ is active}                    \\
    \end{cases}
\end{equation}

With this efficient encoding, the vehicle will then activate the set of final modes $\{M_i^{C[i]} \quad \forall C[i] \neq 0 \}$. Finally, the objective is to maximize the sum of AXIL ratings of each activated mode, while remaining under the current resource budgets $R$ and guaranteeing all dependency constraints in $D$:

\begin{equation} \label{eq:objective}
    \argmax_C \sum_{i=1}^{n_{\text{req}}} X[i,{C[i]]} \quad \text{(note $\forall i, X[i,0] = 0$)}
\end{equation}
$$
    \text{with} \quad \forall k \, \in \, \{1, \, ..., \, r\}, \qquad \sum_{i=1}^{n_{\text{req}}} P_{i, C[i], k} < R[k]
$$
$$
    \text{and} \quad \forall (M_i^j, \, M_k^l) \in E \qquad C[i] \leq j \implies C[k] \leq l
$$

Once we have found the optimal runtime modes, the process is repeated whenever (1) the requested vehicle state $\mathcal{A}_{\text{req}}$ changes, (2) at least one budget increases to allow for higher-end runtime modes, or (3) at least one budget updates to a value below its current resource usage.

\subsection{Problem Complexity}

This problem can be reduced to a variant of the \textit{knapsack} problem \cite{gurski_knapsack_2019} to demonstrate its NP-Hard nature. In the case where applications only define one runtime mode, the problem is equivalent to defining $1$ bin of capacity $R$ with $r$ dimensions. Then, $n_{\text{req}}$ items must be placed in the bin (i.e. runtime modes). Each item is a vector of $r$ resource requirements and has a value defined by its AXIL rating. With the equivalent goal of placing the items with maximized sum of item values in the bin while remaining within the bin capacity on all dimensions, our problem is therefore NP-Hard.

In the case of multiple runtime modes, the complexity increases. Each app can be seen as a set of $m_i$ items. The problem is then to select one item per app. While close to the \textit{multiple-choice knapsack} problem \cite{cacchiani_knapsack_2022}, the problem is different as the items must also follow dependency constraints.

To the best of our knowledge, there is no existing work that solves this problem in an embedded context. In this paper, we propose to use a heuristic to find a solution in polynomial~time.%

\section{Methodology} \label{sec:methodology}

\begin{table}[t!]
    \centering
    \normalsize
    \begin{tabular}{l|l|c}
        \textbf{Setup}   & \textbf{Parameter}               & \textbf{Value} \\
        \hline
        \hline
        \mr{3}{Fixed}    & Population size $n_G$            & 50             \\
                         & Ratio of elitist candidates      & 10\%           \\
                         & Mutation rate                    & 1.15\%         \\
        \hline
        \mr{2}{Variable} & $G_a$ Maximum generations        & 100 to 2000    \\
                         & $G^b$ Auto-stop after stagnation & None to 500    \\
    \end{tabular}
    \caption{Parameters for the genetic algorithm after calibration. See Figure \ref{fig:results_xp} for our choice on the variable parameters. The solver $S$ and heuristic $H$ do not require any parameters.}
    \label{tab:genetics_params}
\end{table}

Our approach enables a dynamic self-organizing orchestration methodology to automatically balance the onboard "experience-to-resource" ratio for best-effort applications.

The purpose of the \textit{Runtime Orchestrator}, illustrated in Figure \ref{fig:arch}, is to manage the selection of runtime modes to achieve the previous optimization goal. It is called when the \textit{State Orchestrator} (out of the scope of this work) changes the combination of requested applications. Once a runtime mode is assigned to each application, the vehicle then launches, stops, or reconfigures all applications to reach the desired state.

However, our proposed approach requires an onboard algorithm to efficiently select modes within the vehicle resources. It also requires a fast response with an optional timeout $t$ if an important application needs to be launched quickly.

In this work, we designed an efficient algorithm in three steps. First, we use a SAT-based solver to obtain a performance baseline. Then, we use a genetic algorithm to improve the ratio between solution quality (i.e. sum of AXIL ratings of running modes) and execution speed. Finally, due to limitations identified in Section \ref{sec:methodology:genetics}, we propose a specialized heuristic to improve the results further and enable onboard implementation. We describe the algorithms in the following paragraphs and compare their performances in Section \ref{sec:results}.

\subsection{SAT Solver} \label{sec:methodology:solver}

First, we implemented our problem in an SMT (SAT-based) solver to obtain a performance baseline \cite{gong_survey_2017}. We use the well-established Z3 solver \cite{bjorner_z_2015} with the objective function and constraints defined in Section \ref{sec:problem}. This approach is guaranteed to return a valid solution if one exists. However, it has an exponential complexity in the worst case. We set a one-hour computation timeout in all scenarios for practical reasons. To model the solution in SAT, we declare $C$ as an unknown vector of variables constrained according to the runtime mode indexes and mode dependencies. Our objective function can be directly translated as a minimization problem, and Z3 will efficiently explore the tree of solutions until the optimal solution is found.

\begin{algorithm}[t!]
    \SetAlgoLined
    \small
    \KwIn{$\mathcal{A}_{\text{req}}$ set of applications, $D$ graph of dependencies, \\ $\quad P$ matrix of resource requirements, $R$ resource budgets}
    \KwSty{Initialization:}{ Vector $C$ with $n_{\text{req}}$ zeros (all apps off)\;}
    \BlankLine

    \While{True}{
    $C^{\text{search}} \gets \emptyset$; \hspace{2.55cm} \textit{Local in-loop candidate}\\
    $S \gets 0$; \hspace{2.2cm} \textit{Local real to hold $C^{\text{search}}$ score}\\
    \BlankLine
    \For{$\{\mathcal{A}_i \in \mathcal{A}_{\text{req}} \quad \forall C[i] \neq 1\}$}{
    \BlankLine
    $C^{\text{new}} \gets C$; \hspace{3.1cm} \textit{Local copy of $C$}\\

    \uIf{$C^{\text{new}}[i] = 0$}{
        $C^{\text{new}}[i] \gets m_i$; \hspace{2.85cm} \textit{Enable $\mathcal{A}_i$}\\
    }
    \Else{
    $C^{\text{new}}[i] \gets C[i] - 1$; \hspace{1.65cm} \textit{Upgrade $\mathcal{A}_i^{C[i]}$}\\
    }

    $C^{\text{new}} \gets \text{bumpDependencies}(C^{\text{new}}, \, D)$; \textit{ Section \ref{sec:methodology:heuristic}}\\

    $s \gets \text{fitness}(\mathcal{A}_i, \, C^{\text{new}}, \, C, \, P)$; \hspace{1.4cm} \textit{Equation \ref{eq:h:fitness}}\\

    \If{$s > 0 \, $ \textbf{and} $\, s > S$}{
        $C^{\text{search}} \gets C^{\text{new}}$; $\quad S \gets s$; \hspace{0.9cm} \textit{Save new best}\\
    }
    \BlankLine
    }
    \BlankLine
    \If{$C^{\text{search}} = \emptyset \, $ \textbf{or} $\, S = 0$}{
        \Return{C} \hspace{1.5cm} \textit{End when no improvement left}
    }
    \BlankLine

    $C \gets C^{\text{search}}$; \hspace{0.45cm} \textit{New best candidate to beat in next loop}\\
    }
    \BlankLine
    \KwOut{Final candidate $C$}
    \caption{Specialized Heuristic $H$}
    \label{algo:heuristic}
\end{algorithm}

\begin{figure*}
    \begin{subfigure}{.2\linewidth}
        \begin{tabular}{>{\small\raggedright}x{2.6cm}|>{\small}x{1.7cm}}
            \textbf{{\LARGE A} \, \, Parameter} & \textbf{Value} \\
            \hline
            Number of apps                      & 20             \\
            Density of deps.                    & 10\%           \\
            Maximum modes                       & 5              \\
            \hline
            Energy Usage                        & 5 to 20\%      \\
            CPU Usage                           & 5 to 20\%      \\
            Memory Usage                        & 5 to 20\%      \\
            Flows per dep.                      & 1 to 5         \\
            Bandwidth per flow                  & 0.1-5Mbps      \\
            \hline
            Budget values for each resource     & 100            \\
        \end{tabular}
    \end{subfigure}
    \hfill
    \begin{subfigure}{.7\linewidth}
        \includegraphics[clip, trim=0.4cm 0.5cm 14cm 0.3cm, width=\linewidth]{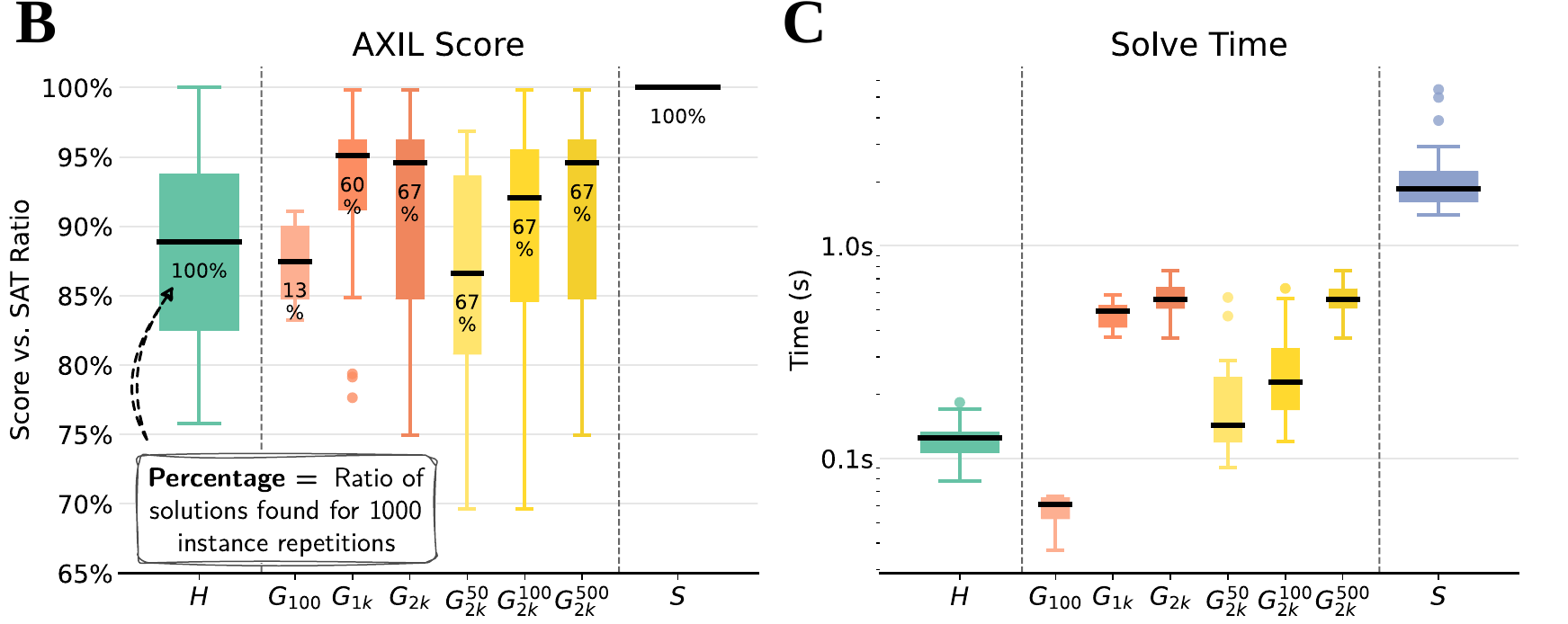}
    \end{subfigure}
    \caption{Simulation results for \textbf{(A)} one particular parameter combination to study different algorithm options. $H$ is our specialized heuristic, $G_a^b$ is a genetic algorithm with $a$ the maximum number of generations and $b$ the number of generations to wait before stopping when the best score hasn't improved (if set), and $S$ is the exact solver based on Z3. We compare the \textbf{(B)} relative score, success rate, and \textbf{(C)} solve time of each algorithm for 1000 random instance repetitions. This figure serves~as an illustration as the overall results greatly depend on the problem size. See Figures \ref{fig:results_move_apps} and \ref{fig:results_mxp} for a general performance assessment.}\label{fig:results_xp}
\end{figure*}

\subsection{Genetic Algorithm} \label{sec:methodology:genetics}

With the aim to embed our methodology onboard, the NP-hard nature of the problem requires a heuristic approach. While several meta-heuristics exist with the potential to solve our problem, we chose to focus on a genetic algorithm $G$ \cite{chu_genetic_1998} for several reasons. First, it is a well-known and widely used approach in the literature. Second, it is capable of evolving in a highly constrained environment, which is required in our case as mode dependencies create a discontinuous search space. Third, the vehicle can stop the algorithm at any time and use the best solution found so far in case of a timeout, which can yield a reasonably performant solution as the algorithm improves quicker in the first iterations. This flexibility cannot be obtained with other meta-heuristics, e.g. simulated annealing.

Defined parameters are summarized in Table \ref{tab:genetics_params}. While the algorithm could be extended with many state-of-the-art variants (e.g. adaptive parameters, diversity preservation), we selected a simple implementation for clarity. The algorithm defines a fixed size population $\{C_1, \dots, C_{n_G}\}$, a mutation rate to avoid overfitting, and a ratio of elitist candidates to improve convergence speed. The notation $G_{a}^{b}$ refers to~a~genetic algorithm with $a$ the maximum number of generations and, if set, $b$ the number of generations to wait before stopping when the score has not improved (or when reaching the timeout~$t$).

For each candidate $C_i$, the fitness function assesses the~score of each candidate and is simply defined as the sum of AXIL scores for each selected mode, i.e. Equation \ref{eq:objective}. The algorithm starts with a random population of $n_G$ candidates, then iteratively generates new candidates by selecting two parents using their scores as selection probability and applying a crossover and mutation operator. The crossover operator defines how two parent genomes should create two children candidates. We chose to select a random pivot point and swap~the modes of the two parents to produce two new candidates. The mutation operator randomly selects a mode and shifts its index by 1. The best candidate is then returned as the final solution.

However, this approach has several limitations. First, the algorithm requires to store several tens of candidates in memory, which is not practical for integration in embedded systems. Second, the algorithm is not guaranteed to return a valid solution, which could hinder on the user experience and bring the vehicle to an undefined state. Finally, as indicated by the $a$ and $b$ parameters, there is no indication on when to stop the algorithm. This is problematic as the algorithm may run for a long time without improving the solution. We discuss these limitations in greater detail with results in Section \ref{sec:results}.

\subsection{Specialized Heuristic} \label{sec:methodology:heuristic}


To overcome the limitations of the genetic algorithm, we propose a specialized heuristic $H$ defined by Algorithm \ref{algo:heuristic} that is guaranteed to return a valid solution although not necessarily optimal. The algorithm starts by selecting a unique candidate $C^0$ where all applications are inactive. Then, the algorithm iteratively 'climbs up' the runtime modes one-by-one, starting from the most degraded modes, using a custom fitness function until no higher runtime mode can be selected.~It stops~when~at least one resource budget $R[i]$ has been reached or 'saturated' (e.g. CPU usage at 100\%, network bandwidth at 100\% capacity for one of the links, or total energy budget reached).

Each iteration follows the following visual metaphor. Using the current candidate $C$, we evaluate the \textit{value-to-resource-usage} ratio improvement of upgrading each application by one mode (if not already at their maximum mode). This can generate up to $n_{\text{req}}$ candidates per iteration. Then, the one with the best fitness score is set as the next iteration reference.

To respect the dependency constraints, this algorithm leverages the property defined in Equation \ref{eq:resources}. When a runtime mode $\mathcal{A}_i^j$ is selected for an application $\mathcal{A}_i$, every mode from other applications that depend on $\mathcal{A}_i$ must also be selected (or any higher mode). This property is enforced by the \textit{bumpDependencies} function which recursively raises the missing mode dependencies until the current candidate respects them all.

Finally, the fitness function compares two candidates and yields an improvement score. This score divides the AXIL difference by a cost function. The cost is defined as the sum of additional resource usages requested by the upgraded mode. If selecting the new mode implies that at least one resource becomes saturated, then the candidate is invalidated in Algorithm \ref{algo:heuristic}. Additionally, the cost is designed to be lower when a resource is filled in its lower capacity to encourage the progressive allocation of resources. Hence, we use the integral of the resource usage difference. The cost function for one resource $k$ with budget $b$ used from $u$ to $u'$ is defined as:

\vspace{-.2cm}
\begin{equation*}
    f(u, u', b) = \begin{cases}
        0          & \text{if} \quad u' > b                                       \\
        u'^2 - u^2 & \text{otherwise} \quad (= \frac{1}{2} \int_{u}^{u'} x \, dx)
    \end{cases}
\end{equation*}

Then, the fitness function sums the cost of each resource to obtain the final score or returns 0 if one resource is saturated:

\vspace{-.2cm}
\begin{equation} \label{eq:h:fitness}
    \text{fitness}(\mathcal{A}_i, \, C^{\text{new}}, \, C, P) = \begin{cases}
        0                                                    & \text{if} \,\, 0 \in K \\
        \frac{{X[i,C^{\text{new}}[i]] - X[i,C[i]]}}{\sum{K}} & \text{otherwise}
    \end{cases}
\end{equation}
$$
    \text{with} \quad K = \{f(M_i^{C^{\text{new}}[i]}[k], \, M_i^{C[i]}[k], \, R[k]) \quad {\scriptstyle \forall k = 1, \dots, r} \}
$$

The algorithm stops when no higher runtime modes can be selected once resources are used (i.e. when all scores are $0$).


\subsection{Algorithmic Complexity}
Contrary to the genetic algorithm $G$ which stores and maintains many candidates, our heuristic $H$ only stores three candidate solutions during the search, two being temporary.

Our proposed heuristic is of polynomial time complexity. Each iteration must evaluate $n_{\text{req}}$ candidates, each dependent on a fitness function that sums $r$ costs. In the worst case, the algorithm may need to upgrade every available mode one by one. Hence, the overall complexity is $\mathcal{O}(m \times n_{\text{req}} \times r)$. However, due to its efficient search strategy, the algorithm is fast in practice as demonstrated by our results in Section \ref{sec:results}.

Finally, the guarantee that $H$ will always return a valid solution is due to its search direction enabled by Equation \ref{eq:resources}. The first iteration activates one of the lowest runtime modes with the best \textit{AXIL-to-resource-usage} ratio. As long~as there is at least one mode that does not immediately saturate a resource, this heuristic will certainly activate at~least one application. Then, by iteratively upgrading modes, the algorithm slowly fills the resources. We rely on the extensive simulation results presented in Section \ref{sec:results} to discuss the overall performance.

In practice, this time complexity may limit onboard scalability to a few tens of applications. However, further evaluations in a realistic embedded environment are required to assess the practical usability of our approach, out of the scope of this purely algorithmic study. We believe this remains within the probable problem sizes managed by future vehicles.

\section{Simulation setup} \label{sec:simulation}

To evaluate the performance of each algorithm ($S$, $G$, and $H$) and assess their embedded feasibility, we designed a simulation setup with randomly generated applications, resource requirements, mode dependencies, and network flows. The list of parameters required to generate a random instance along with selected values for our first study is shown in Figure~\ref{fig:results_xp}A. Values were chosen to represent what we believe to be a median use case after internal discussions within Stellantis.

The problem instance is generated as follows. First, we model a fixed physical E/E topology as a star network of 5 ECUs connected by Ethernet links of 100Mbps. The central ECU could be seen as the main onboard HPC, and the others as zonal (ZCU) or communications (CCU) ECUs. Note that our approach is independent of the topology type. Then, we generate a random number of $n$ applications to represent an app store. Each application is randomly assigned to one of the ECUs, which we assume is provided by an external onboard service orchestrator out of the scope of this work. Additionally, we generate a random dependency graph with a target density by iteratively adding edges randomly to a graph up to a given target density, and removing one edge per cycle if any appears.

Each app $\mathcal{A}_i$ has a random number of $m_i$ runtime modes below a maximum value $m_{\text{max}}$. For each dependency edge in the app-level graph, we generate a new mode-level dependency graph $D$. We generate a random number of edges between modes of the two apps as long as they do not cross. Then, for each edge in $D$, we generate a random number of network flows with random bandwidth requirements. Finally, each mode $\mathcal{A}_i^j$ is attributed a random resource requirement $M_i^j [k]$ for each resource $k$, i.e. CPU, RAM, and energy consumption in this work. Note that values are generated within the bounds set in Figure \ref{fig:results_xp}A. They also respect all constraints defined in Section \ref{sec:problem}. See Figure \ref{fig:instance} for an example of AXIL ratings generated for a particular instance with values in Figure \ref{fig:results_xp}A.

Finally, each value is stored in a 3D resource matrix $P$ of $r$ layers, i.e. made of 2D matrices for each resource. In total, there are $r=21$ layers, namely the global energy usage requirement, CPU and memory requirements for each ECU, and network bandwidth usage for each direction of the 5 Ethernet cables. Bandwidth requirements for each flow are added in each link layer that is part of the packet route.

Each algorithm receives a given randomly generated problem instance as input and returns both their best candidate solution $C$ and total solve time. The latter includes initialization steps such as creating assertion rules in the solver or population generation in $G$. $H$ only sets a vector of $n_{\text{req}}$ zeros.

\section{Results} \label{sec:results}

\begin{figure}[t!]
    \centering
    \includegraphics[clip, trim=0.4cm 0.5cm 0.6cm 0.4cm, width=\linewidth]{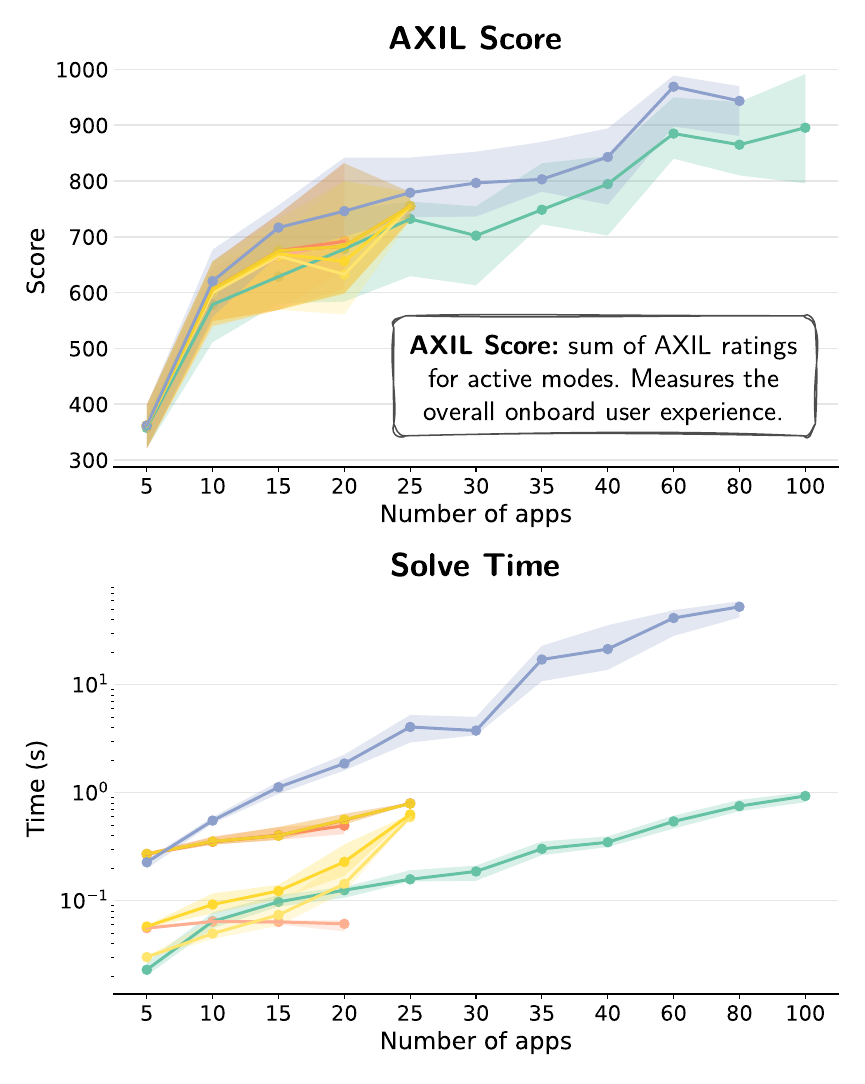}
    \caption{Performance evaluation by moving the number of applications requested to launch from 5 to 100, with other parameters kept identical as Figure \ref{fig:results_xp}A. Each color follows the legend from Figure \ref{fig:results_xp}B. Lines represent the median value of 100 repetitions and corresponding areas the Q1 to Q3 quartiles.}
    \label{fig:results_move_apps}
\end{figure}

\begin{figure*}[t!]
    \centering
    \includegraphics[clip, trim=0cm 0cm 0cm 0cm, width=.93\linewidth]{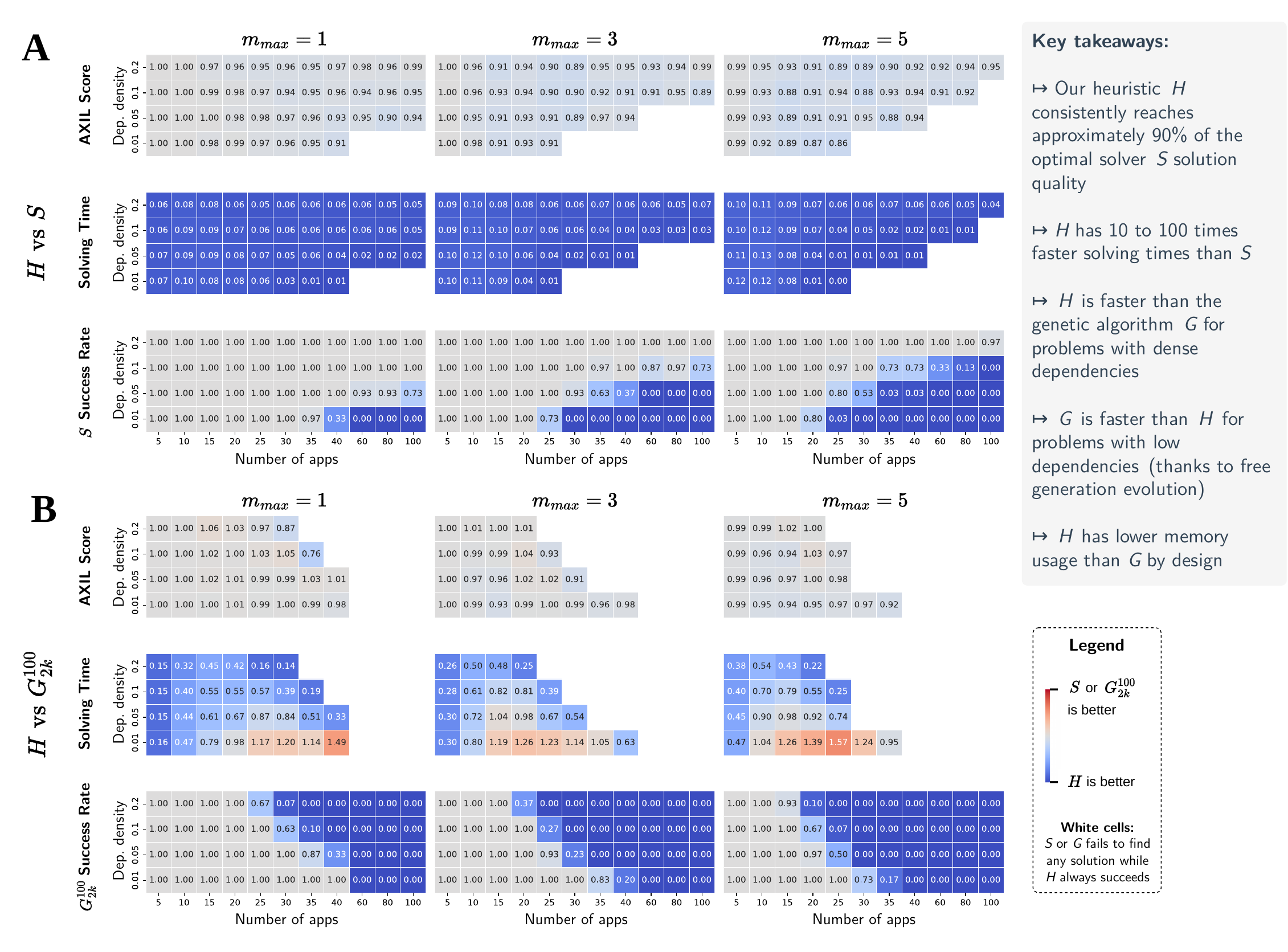}
    \caption{Multi-parameter performance comparison of our proposed heuristic algorithm $H$ with the \textbf{(A)} ILP solver $S$ and \textbf{(B)} genetic algorithm $G_{2k}^{100}$ chosen manually as the best competitor variant. Each cell represents the ratio of the median value from 30 instance repetitions on performance, solve time, or success rate. When a cell is empty, either $S$ or $G$ failed to find a~solution for all repetitions while $H$ always finds all solutions. Hence, the success rate corresponds to $S$ (1h timeout) or $G$.}
    \label{fig:results_mxp}
\end{figure*}

The simulation results greatly depend on the problem instance parameters. We first present the results for a single set of parameters to understand the main trends in Figure \ref{fig:results_xp}. Then, Figure \ref{fig:results_move_apps} shows the evolution of performance depending on the number of requested applications, which is the main parameter that can change dynamically. Finally, we present a general performance assessment through a heatmap in Figure \ref{fig:results_mxp} for all parameter combinations. A clear trend will be observed.

With 1000 instances generated using parameter values in Figure \ref{fig:results_xp}A, Figure \ref{fig:results_xp}B shows the distribution of the total AXIL score for each algorithm relative to the value found by the solver (set as $100\%$). Figure \ref{fig:results_xp}C shows the solving time of each algorithm. Since algorithm $G$ does not provide any stop condition, we studied 6 variants with different fixed stop points or once the score does not evolve after a number of iterations.

We observe that both the performance and solve times are similar between our heuristic $H$ and $G$. However, while $G_{100}$ runs faster than $H$, it only finds a valid solution in $13\%$ of all instances. All other $G$ variants take slightly more time to return a solution than $H$ and have a success rate of $60$ to $67\%$, while $H$ always returns a valid solution with relative good performance. While there is a clear advantage of using $H$ in this scenario (parameters set in Figure \ref{fig:results_xp}A), we then extended the comparison with other parameter combinations.

Since the problem is defined by many parameters, we chose to only vary the number of applications, modes, and density of applications. We believe this to be a reasonable simplification as all other parameters correspond to usually fixed app-level resource requirements. Figure \ref{fig:results_move_apps} focuses on studying the sensitivity of results on the number of applications requested to launch. We believe that this parameter may vary the most at runtime. With 100 instances per problem size, we observe that the AXIL scores found by all algorithms are very close, which indicates good performance for both $G$ and $H$. However, all variants of $G$ fail after a \textit{frontier} of complexity while $H$ always succeeds. Note that the 1h timeout for $S$ is reached at the last point for all instances. After this point, the performance of $H$ cannot be compared, although it can still be used in practice.

Finally, Figure \ref{fig:results_mxp} shows a heatmap of the performance of each algorithm for all parameter combinations. Each cell represents the ratio of the median value from 30 instance repetitions on performance, solve time, and success rate. When a cell is empty, either $S$ or $G$ failed to find a solution for all repetitions while $H$ always finds a solution for every instance. Hence, the success rate corresponds to $S$ or $G$. Both algorithms fail after reaching their own frontier of problem complexity.

We observe that $H$ is always the best performing algorithm in terms of performance and solve time. $G$ is faster than $H$ for small problem sizes, mainly with fewer mode dependencies, although the absolute computing time could be negligible for both cases. This is due to the fact that $G$ is a genetic algorithm that relies on a population of candidates to evolve towards a better solution using random crossover operations. When dealing with many mode dependencies, the search space becomes discontinuous (like holes in cheese) and the population struggles to generate valid children. This is also why $G$ is faster than $H$ for small problem sizes, as crossovers and mutations are very effective under low constraints.

While the problem could potentially be solved by more efficient approaches than our proposed heuristic, the proposed algorithm in this work is both fast and always returns a valid solution. This is a key requirement for onboard runtime decisions, as the algorithm must always return a valid solution within a reasonable time frame. In case where $H$ fails to find a solution within the timeout $t$, we can use its latest candidate solution as a fallback and optionally continue the search later.

\section{Discussion}

By introducing AXIL, we propose a way of expressing a launch priority based on the user's perceived quality of experience for each runtime mode. Our approach aims to solve the problem of orchestrating a high number of applications in a highly constrained embedded system. This concept led us to build a resource-based model of the onboard vehicle state, which in turn enables optimization algorithms to maximize the total onboard AXIL score given the resource constraints.

This approach can be particularly useful in two situations. First, the user and background services can be requested to launch without limit, and our algorithm selects the best experience-to-resource ratio to accommodate for the current global onboard state. This reduces the need for state management which is not mandatory for non safety-critical applications thereby reducing engineering effort. It also increases the number of active applications by potentially executing more of them in slightly degraded modes. Second, AXIL ratings can change depending on the user preferences and subscriptions, which impacts the autonomous runtime decisions and tailors the experience to the user. Hence, SLAs can be respected by tweaking AXIL ratings based on the user profile or contract.

Our heuristic starts with all applications disabled and gradually improves runtime modes based on an AXIL and resource-based cost function. This approach offers great potential flexibility in an industrial setting, allowing automakers to initiate the algorithm with a few pre-selected modes thereby guaranteeing their activation. They are guided by marketing strategies, user preferences, state management constraints, safety-critical prerequisites, or V2X service priorities.

Additionally, while this work focuses on onboard resources, our model allows for a flexible incorporation of various other resource types to enhance the overall system. For instance, edge computing resources near the current vehicle location or internal peripherals on each ECU can be added as additional resources. Finally, automakers can add a "digital eco-mode" onboard by considering energy consumption as an auxiliary resource, empowering users to select their preferred trade-off between user experience and environmental impact. This extended metric could include network infrastructure and data center usage to evaluate each runtime mode's total impact \cite{yan_modeling_2019}.

Our proposed model currently has operational limitations. First, it is centralized and requires a global vehicle state to be available at all times. If this condition fails (e.g. link or software component failure), the vehicle may need to fall back to a pre-defined default behavior. Second, the cost of calling the algorithm may not be insignificant and should be considered when integrating it into the vehicle. Future strategies may optimize the selection of execution times based on resource availability dynamics to reduce the overall onboard overhead.

While the algorithm's complexity necessitates careful consideration of response times and resource dynamics, its scalability and autonomous optimization offer significant benefits to automakers. This dynamic methodology not only reduces hardware requirements but also facilitates personalized user experiences and a robust app ecosystem, marking a significant paradigm shift from traditional static OTA updates.

\section{Conclusion}

We have presented a methodology for orchestrating non-safety-critical onboard applications based on the available vehicle resources to cover the dynamic future automotive use cases. Once developers define several degraded modes and dependencies for each application, our approach then selects the highest possible modes without exceeding the current onboard resources while maximizing the overall user experience. Our proposed heuristic achieves a performance-to-speed ratio that enables onboard adjustments at runtime while also guaranteeing to always return a valid solution, contrary to traditional approaches like using a SAT solver or a genetic algorithm. Finally, the vehicle invokes our algorithm on context or resource availability changes and dynamically applies the calculated runtime modes for a continuous user~experience.


Limitations of our approach concern its scalability in a real-time environment, as its polynomial time complexity likely limits its practical usability to a few tens of concurrent onboard applications. While we believe this remains within the probable problem sizes managed by future vehicles, a realistic performance assessment in an industrial context including practical optimizations would appear to be beneficial.

Future work will aim to (1) evaluate the industrial feasibility of our proposal using a physical test bench with realistic use cases and embedded resources, (2) evaluate our algorithm's potential to influence on the onboard energy consumption \cite{weber_energy-optimized_2020}, \changed{(3) extend our work to better support dynamic AXIL ratings, resource requests by applications, and environment changes,}\removed{ (3) dynamically optimize mode selection based on a future timeline of predicted events and context,} and (4) discuss practical implementation strategies such~as operational constraints and AXIL definition~methodologies.



\bibliographystyle{IEEEtran}
\bibliography{references}

\vspace{-.85cm}

\begin{IEEEbiography}[
        {\includegraphics[width=1in,height=1.25in,clip,keepaspectratio]{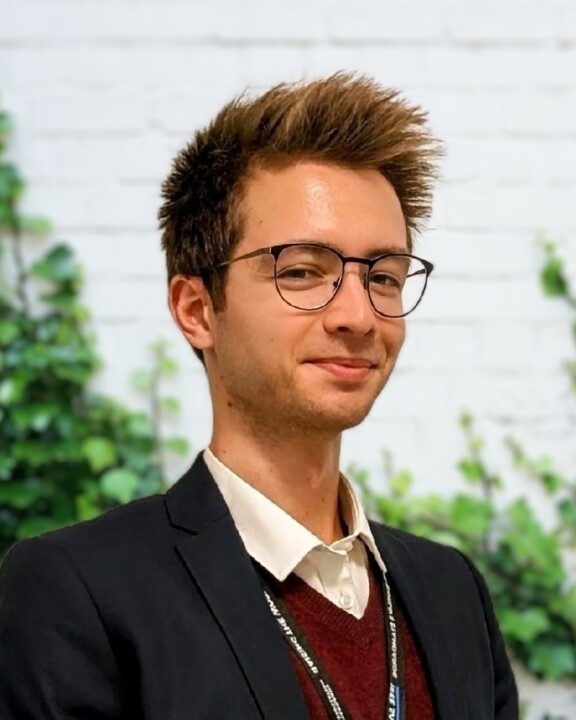}}]{Pierre Laclau}
    is an industrial PhD student associated with Stellantis and the Heudiasyc laboratory, CNRS, Université de Technologie de Compiègne (UTC), France. He has a double degree of both Computer Science Engineering (with a specialization on real-time, embedded, and complex systems) and a Master of Science (specialized on autonomous and intelligent systems) both validated at UTC. His research interests lie in multi-agent and complex systems architecture design applied to the development of next generation E/E and software architectures.
\end{IEEEbiography}

\vspace{-.85cm}

\begin{IEEEbiography}[
        {\includegraphics[width=1in,height=1.25in,clip,keepaspectratio]{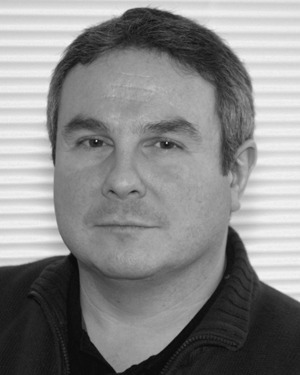}}]{Stéphane Bonnet}
    received the M.Sc. degree in computer science and the Ph.D. degree in control engineering from the Université de Technologie de Compiègne, Compiègne, France. He is currently a Research Engineer at this same institution and has contributed to several research projects, covering a wide array of related fields in embedded systems engineering, robotics, and communications. His current main interests are autonomous vehicles, related vehicle-to-vehicle and vehicle-to infrastructure communication systems, and their impact on smart cities.
\end{IEEEbiography}

\vspace{-.85cm}

\begin{IEEEbiography}[
        {\includegraphics[width=1in,height=1.25in,clip,keepaspectratio]{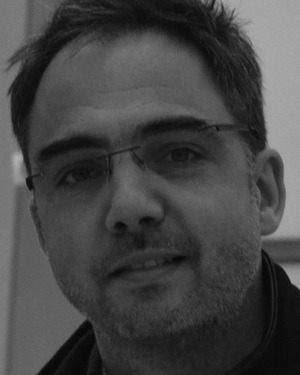}}]{Bertrand Ducourthial}
    Bertrand Ducourthial received the Ph.D. degree in computer science from the University of Paris-Sud, Orsay, France, in 1999. He joined the Heudiasyc Laboratory, Université de Technologie de Compiègne, Compiègne,  France, where he became a Full Professor in 2010. His research deals with dynamic networks, with applications in vehicular ad hoc networks.
\end{IEEEbiography}

\vspace{-.85cm}

\begin{IEEEbiography}[
        {\includegraphics[width=1in,height=1.25in,clip,keepaspectratio]{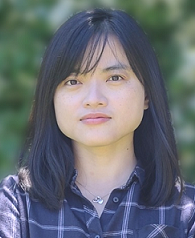}}]{Trista Lin}
    is a software architect at Stellantis since 2018. She is responsible for TCP/IP architecture design and protocol deployment. She holds a PhD degree in computer science from INSA Lyon (National Institute of Applied Sciences of Lyon), France, and a BS in mathematics and communications engineering from National Tsing Hua University, Taiwan. Her research interests lie in IT solution adaptation for cars towards software-defined architectures and services.
\end{IEEEbiography}

\vspace{-.85cm}

\begin{IEEEbiography}[
        {\includegraphics[width=1in,height=1.25in,clip,keepaspectratio]{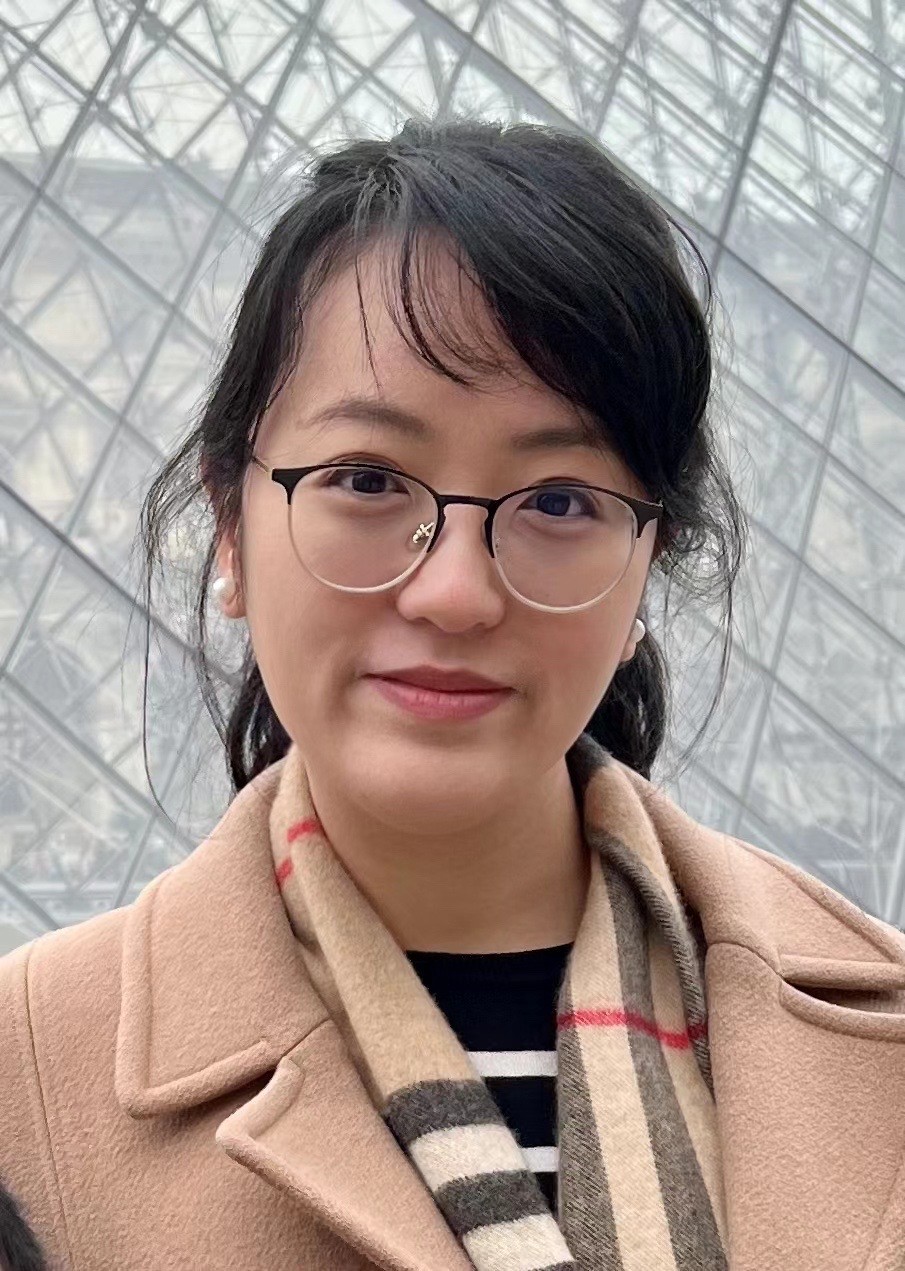}}]{Xiaoting Li}
    is an In-Vehicle Network engineer at Stellantis since 2018. She holds a Ph.D. degree in Computer Science from the National Polytechnic Institute of Toulouse, France in 2013. Prior to joining Stellantis, she worked in high education sector as an associate professor in real-time network and system domain. She is currently working on In-Vehicle Network technologies, with a particular focus on Ethernet and associated protocols such as AVB/TSN and Software-Defined Vehicle for the next generation of E/E automotive architecture.
\end{IEEEbiography}

\end{document}